\documentstyle[twocolumn,pra,aps,epsfig]{revtex}
\def\be{\begin{equation}}
\def\e#1{\label{#1}\end{equation}}
\def\bea{\begin{eqnarray}}
\def\ea#1{\label{#1}\end{eqnarray}}
\def\r#1{(\ref{#1})}
\def\bem#1{\begin{mathletters}\label{#1}}
\def\eml{\end{mathletters}}
\def\df{\Delta\varphi}

\begin{document}
\draft
\title{Zeno and anti-Zeno effects for photon polarization dephasing}
\author{A. G. Kofman$^{1}$, G. 
Kurizki$^{1}$ and T. Opatrn\'{y}$^{1,2}$}
\address{
$^1$ Department of Chemical Physics, Weizmann Institute of Science,
761~00 Rehovot, Israel \\
$^{2}$ 
Theoretisch Physikalisches Institut, Friedrich-Schiller 
Universit\"{a}t, Max-Wien-Platz~1, D-0743 Jena, Germany, 
\\ and
Department of Theoretical
Physics, Palack\'{y} University, Svobody 26, CZ-77146 Olomouc, Czech 
Republic
}

\maketitle
\begin{abstract}
We discuss a simple, experimentally feasible scheme,
which elucidates the principles of controlling (``engineering'')  
the reservoir spectrum and
the spectral broadening incurred by repeated measurements. 
This control can yield either the inhibition (Zeno effect) or the 
acceleration (anti-Zeno effect) of the quasi-exponential decay of 
the observed state by means of frequent measurements. 
In the discussed scheme, a photon is bouncing back and forth between 
two perfect mirrors, each time passing a polarization rotator. 
The horizontal and vertical polarizations can be viewed as analogs of 
an excited and a ground state of a two level system (TLS).
A polarization beam splitter and an absorber for the vertically 
polarized photon are inserted between the mirrors, and effect
measurements of the polarization.
The polarization angle acquired in the electrooptic polarization 
rotator can fluctuate randomly, e.g., via noisy modulation. 
In the absence of an absorber the polarization randomization 
corresponds 
to TLS decay into an {\em infinite-temperature} {\em reservoir}. 
The non-Markovian nature of the decay stems from the many round-trips 
required for the randomization.
We consider the influence of the polarization 
measurements by the absorber on this non-Markovian decay, and develop
a theory of the Zeno and anti-Zeno effects in this system.

\end{abstract}


\section{Introduction}

The quantum Zeno effect (QZE) is the striking prediction that the 
decay of {\em any\/} unstable quantum state can be inhibited by 
sufficiently frequent observations (measurements) 
\cite{kha68,fon73,mis77}. 
The QZE has been experimentally tested \cite{ita90} and primarily 
analyzed for {\em two coupled states} \cite{joo84,coo88,fre91} (with 
few exceptions \cite{wil97,ela00}). 
Yet the consensus opinion has upheld the QZE as a {\em general\/} 
feature of quantum mechanics which should lead, e.g., to the 
inhibition of radioactive or radiative decay \cite{sak94,sch98}. 
The claim of QZE generality has rested on the assumption that 
successive observations can, {\em in principle}, always be made at 
time intervals too short for the system to change appreciably. 
However, this assumption and the 
generality of the QZE have scarcely been investigated. 

We have now shown \cite{Nature} that this assumption is basically
incorrect and that the QZE does not hold generally, but only in a 
restricted class of systems.
The main implications of our theory are: 
(i) The QZE is principally unattainable in 
radiative or radioactive decay, because the required measurement 
rates may cause the system to disintegrate (via the production of new
particles).
(ii) Decay {\em acceleration\/} by frequent measurements (the 
anti-Zeno effect - AZE) \cite{Nature,Abraham,Gil,lew00} is possible
for essentially {\em any} decay process, and is thus much more 
ubiquitous than its inhibition (the QZE).
These findings stem from the {\em universal} result 
\cite{Nature}, whereby the modification of the decay rate by frequent 
measurements is determined by the convolution (overlap) of two
functions: (a) the measurement-induced
spectral broadening ({\em energy spread}), which is 
proportional to the rate of measurements, in accordance with 
the time-energy uncertainty relation; (b) the {\em spectrum
of the reservoir} (bath) to which the decaying state is coupled. 
The QZE or AZE correspond to the measurement-induced spread being much
broader or narrower than the reservoir spectral width, respectively.
The {\em non-Markovian nature of any physical decay process}, 
associated with the {\em finite\/} spectral width (and, 
correspondingly, nonzero memory time) of the reservoir, is the 
essential property allowing the modification of the decay by means 
of frequent measurements, be it the QZE or the AZE.

The universal formula \cite{Nature} was obtained on the basis of the
projection postulate.
In reality there can be many different measurement schemes, which
can be classified as direct, when the initial state itself is 
measured \cite{ita90,coo88,fre91,ela00}, and indirect 
\cite{sch98,per80,pan96}, when the final state(s) are measured.
Whereas direct measurements should be nondestructive, i.e., conserving
the population of the measured state, indirect measurements can be 
either nondestructive or destructive.
This difference can affect the dynamics for large times, when the
initial-state population significantly differs from 1.
However, for short times, when this population is close to one, 
formula obtained in \cite{Nature} should hold for the both types of
measurements.

In this paper we discuss a simple, experimentally feasible scheme,
which elucidates the principles of controlling (``engineering'')  
the reservoir spectrum and
the spectral broadening incurred by repeated measurements. 
This control can yield either the inhibition (QZE) or the 
acceleration (AZE) of the quasi-exponential (non-Markovian) decay of 
the observed state by means of frequent measurements. 
In the present scheme, the pertinent observable is photon
polarization, constituting the optical analog of a two-level system
\cite{per80}.
The photon is injected into the setup via a fast gate, which first
reflects and then transmits horizontal polarization.
The injected photon is bouncing back and forth between two perfect 
mirrors, each time
passing a polarization rotator (Fig. \ref{fig1}) \cite{Kwiat}. 
A polarization beam splitter (PBS), which reflects a vertically 
polarized photon and transmits a horizontally polarized one, as well 
as an absorber, are inserted between the mirrors. 
If the path of the vertically polarized photon is completely blocked 
by the absorber, then it realizes a discrete measurement (projection) 
at each passage: the photon is either lost or its state is projected 
onto the horizontal polarization. 
This situation has been used to demonstrate an interaction-free
measurement \cite{eli93,kwi95,kwi99}.
If, on the other hand, the absorber is partially transparent to 
vertically polarized photons, then it realizes an imperfect
measurement \cite{mil88,per90}.

The polarization angle acquired in the electrooptic polarization 
rotator can 
fluctuate randomly, e.g., via noisy modulation of a Pockels cell. 
In the absence of the absorber, the polarization, after {\em many 
round-trips,\/} then becomes random and the probability of finding 
any particular polarization tends to 1/2. Taking the horizontal 
and vertical polarizations as analogs of an excited and a ground 
state of a two level system, this polarization randomization 
corresponds to decay into an {\em infinite-temperature} {\em 
reservoir}. 
The non-Markovian nature of the decay stems from the many round-trips 
required for the polarization 
to change randomly, which implies a {\em long memory time\/}. 
Our goal is to consider the influence of the polarization 
measurements by the absorber on this non-Markovian decay, and derive 
the conditions of the QZE and AZE in this system.
It should be noted that if the source of injected photons is a laser,
governed by quasiclassical photon statistics, then the results of
polarization decay can be interpreted classically.
However, the concept of {\em the Zeno effect is valid in classical
electromagnetism\/}, as pointed out by Peres \cite{per80}.

In Sec.~\ref{Sec-Model} we present the physical model and its 
features in the limits of perfect absorption or fixed (rather than
random) rotation angles. 
In Sec.~\ref{Sec-Analysis} the master equations for its general 
analysis are derived (any absorption and rotation). 
Sec.~\ref{Sec-continuous} is devoted to continuous dephasing, i.e.,
the limit of small, highly correlated random rotations and weak 
absorption (ineffective measurements).
Sections~\ref{Sec-discrete} and \ref{VI} deal with discrete dephasing 
(namely, correlated and anticorrelated phase jumps) and arbitrary
absorption (effective and ineffective measurements).
Conditions for the QZE and AZE are derived.
In particular, in Sec.~\ref{Sec-discrete} a general theory is
developed for the most interesting case of {\em small rotation 
angles}, whereas in Sec. \ref{VI} a simple non-Markovian model for 
random rotations of an {\em arbitrary size} is studied analytically 
and numerically.
The conclusions are given in Sec.~\ref{Sec-concl}.
Appendices \ref{A}, \ref{B} and \ref{C} give the details of the 
general analysis, the small and the arbitrary-size phase-jump 
analysis, respectively.


\section{Model description}
\label{Sec-Model}
\subsection{The setup}

\begin{figure}[htb]
\centerline{
\begin{tabular}{cc}
\epsfig{file=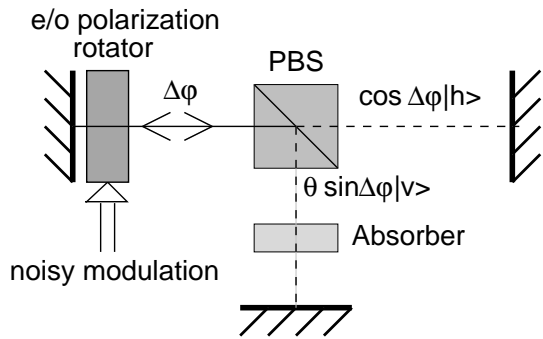,width=2.8in}
\end{tabular}
}
\protect\caption{Setup for controlling the polarization decay of a
photon bouncing between the mirrors. Measurements are effected by a
polarization beam-splitter (PBS) and an absorber with variable
transparency $\theta$. 
The ``reservoir'' into which the polarization decays is realized upon 
modulating a Pockels cell (which rotates the polarization by $\df$) 
by a field with controllable noise properties.}
\label{fig1}
\end{figure}

Consider the setup in Fig. \ref{fig1} \cite{Kwiat}.
A horizontally polarized photon, denoted as $|h\rangle$, enters the 
setup via a fast gate (not shown), which changes from being totally 
reflective to totally transparent to $|h\rangle$ on a ns scale.
The polarization rotator between the two highly reflecting mirrors  
causes fast (ns-scale) rotation of the photon polarization by means 
of a Pockels cell or another electrooptic element.
The $|h\rangle$ photon, transmitted by the PBS, bounces between the
mirrors, while a vertically polarized photon, denoted as $|v\rangle$,
which is reflected by the PBS, is blocked by an absorber, which can 
be made {\em partially transparent} with transmissivity $\theta$.
(Alternatively, one can use a perfect absorber and a PBS which is
partially transparent for the vertically polarized photon 
$|v\rangle$.)

\subsection{QZE for constant rotation angle and complete absorption}

When the path of $|v\rangle$ is open, the probability of finding the 
photon with horizontal polarization after $n$ round trips is
\begin{eqnarray}
  P_{h}(n) = \cos ^{2}(n\Delta \varphi), 
\label{4}
\end{eqnarray}
assuming a fixed $\Delta \varphi$, the rotation angle of the 
polarizer at each passage: the polarization oscillates between the 
vertical and horizontal states, analogously to a Rabi oscillation.
If, on the other hand, the path of  $|v\rangle$  is completely 
blocked by a perfect absorber, this absorber realizes an impulsive 
measurement (IM), i.e. a projection, at each passage:
the photon is  either lost or its state is
projected onto the $|h\rangle$ state. 
With the completely blocked  $|v\rangle$ path the probability of
finding $|h\rangle$ decays as
\begin{eqnarray}
  \label{expdecay}
  P_{h}(n) = \cos^{2n}(\Delta \varphi), 
\end{eqnarray}
where $n$ is the number of the round-trips.
With $\Delta \varphi$ $\ll$ 1 the {\em effective exponential decay is 
much slower than the Rabi oscillation}. 
This constitutes an example of the QZE by IMs 
\cite{ita90,coo88,Nature,Abraham,kwi99}, which has been introduced as
a demonstration of interaction-free measurements 
\cite{eli93,kwi95,kwi99}.

\subsection{Arbitrary absorption and rotation}

To extend the above results to the case of partial absorption and
arbitrary rotation angle $\Delta\varphi_k$ ($k=1,\dots$), we write 
down the obvious dynamical equations for the horizontal and vertical 
components of the photon field envelope (wave function), 
$\epsilon_{h}$ and $\epsilon_{v}$, respectively, 
\bea
&\dot{\epsilon}&_{h}=-\mu(t)\epsilon_{v},\nonumber\\
&\dot{\epsilon}&_{v}=\mu(t)\epsilon_{h}-
\frac{\Gamma(t)}{2}\epsilon_{v}.
\ea{1}
Here $\mu(t)=\dot{\varphi}(t)$, $\varphi(t)$ is the total angle
accumulated by consecutive $\Delta\varphi_k$ between the
polarization vector and the vertical axis of the polarizer,
and $\Gamma(t)$ is the photon absorption rate in each passage. 
The rates of polarization rotation and absorption, $\mu(t)$ and 
$\Gamma(t)$, are time-periodic functions, whose period is
the round-trip time $\tau_{\rm r}$. 
They do not temporally overlap in the setup of Fig. \ref{fig1}, i.e., 
there exists $\tau_1$ ($\tau_1<\tau_{\rm r}$) such that $\mu(t)$
vanishes for $\tau_1\le t\le\tau_{\rm r}$ while $\Gamma(t)$ vanishes 
for $0\le t\le\tau_1$. 
Equation~(\ref{1}) is equivalent to the Schr\"{o}dinger equation for 
an {\em open two-level system\/} (TLS) {\em interacting with a 
field\/} $\mu(t)$. 

The general solution of Eq. (\ref{1}) at $t=n\tau_{\rm r}$ is
\be
\left(\begin{array}{c}\epsilon_h(t)\\ 
\epsilon_v(t)\end{array}\right)=\left[\prod_{k=1}^n
\left(\begin{array}{lr}\cos\Delta\varphi_k&-\sin\Delta\varphi_k\\
\theta\sin\Delta\varphi_k&\theta\cos\Delta\varphi_k\end{array}\right)
\right]
\left(\begin{array}{c}\epsilon_h(0)\\ 
\epsilon_v(0)\end{array}\right),
\e{7}
where
\be
\theta=\exp\left[- \frac{1}{2}\int_0^{\tau_{\rm r}}\Gamma(t)dt\right]
\e{3}
is the amplitude transmission coefficient of the absorber.
Within the above constraints $\mu(t)$ and $\Gamma(t)$ are {\em
arbitrary}. 
Most of the subsequent analysis will be dedicated to {\em random} 
$\mu(t)$ caused by noisy modulation of the electro-optic rotator.

\subsection{QZE for fixed rotation angle and incomplete absorption}

Assuming again a {\em fixed} $\Delta\varphi$ (all $\Delta\varphi_k$
being equal), one obtains from \r{7} the following time-dependent
probability for the photon to keep its initial horizontal polarization
($\epsilon_{h}(0)=1$, $\epsilon_{v}(0)=0$)
\bea 
&&P_{h}(t=n\tau_{\rm r})=\epsilon_{h}^2(t)\nonumber\\
&&=\{[\lambda_1^n(\cos\Delta\varphi-\lambda_2)+\lambda_2^n
(\lambda_1-\cos\Delta\varphi)]/D\}^2,
\ea{2}
where $D=\sqrt{(1+\theta)^2\cos^2\Delta\varphi-4\theta}$ and 
$\lambda_{1,2}=[(1+\theta)\cos\Delta\varphi\pm D]/2$. 
For complete transparency, $\theta=1$, Eq. \r{2} reduces to the 
result \r{4} and for complete absorption, $\theta=0$, to the result
\r{expdecay}.

For small phase jumps $\Delta\varphi$ and sufficiently strong 
absorption, $(\Delta\varphi)^2\ll(1-\theta)^2$,
the photon polarization decay is approximately exponential
\bea
P_{h}(t=n\tau_{\rm r})=&\exp&\left[-\frac{(1+\theta)
(\Delta\varphi)^2}{(1-\theta)\tau_{\rm r}}t\right].
\ea{5}
It is intuitively clear that the effective rate of measurements $\nu$ 
increases with the quantity $1-\theta$, which plays the role of the 
effectiveness of the measurements. 
This will be rigorously confirmed in Sec. \ref{Sec-discrete}.
Correspondingly, the decay rate in Eq. \r{5} decreases with the 
increase of $1-\theta$, thus demonstrating the QZE in the case of
fixed $\Delta\varphi$. 


\section{General analysis of polarization dephasing: Master equations}
\label{Sec-Analysis}

In the ensuing analysis we assume that the {\em rotation angle\/} 
$\Delta \varphi$ {\em is not fixed but fluctuates randomly\/}, e.g., 
due to the modulation of the Pockels cell rotator by a {\em noisy 
control field}.
Thus, after many round trips the polarization of
the photon becomes random and the probability of finding any 
particular polarization tends to 1/2. 
Taking the horizontal and vertical polarizations as analogs of the
excited and ground states of a two level system (TLS), {\em this 
polarization randomization corresponds to decay into an 
infinite-temperature reservoir.\/} 

In what follows, we shall write down the most general equations
describing the influence 
of polarization projection measurements on such decay.
The polarization tensor (or the density matrix for the TLS) obeys the
equation
\be
\dot{Q}=[A(t)+C\mu(t)]Q.
\e{6}
Here $Q=(P_{h},P_{v},u)^\dagger$, $P_{v}(t=n\tau_r)$ $\equiv$ 
$\epsilon_{v}^2(t)$ is the
probability for the photon to have the vertical polarization,
$u=2\epsilon_{v}\epsilon_{h}$ is the coherence, in TLS terms, 
or the first Stokes parameter \cite{lan75}, and
\bea
A(t)=\left(\begin{array}{lcr}0&0&0\\
0&-\Gamma(t)&0\\
0&0&-\frac{\Gamma(t)}{2}\end{array}\right),\quad
C=\left(\begin{array}{lcr}0&0&-1\\
0&0&1\\2&-2&0\end{array}\right).
\ea{8}
Equation (\ref{6}) can be reduced to an equation for the average 
quantity $\bar{Q}(t)$, involving an expansion in cumulants of 
$C\mu(t)$\cite{ter74,kof90}. 
As shown in Appendix \ref{A}, by truncating the cumulant expansion at 
the second order, one obtains the following {\em non-Markovian\/} 
differential master equations for the average polarization 
probabilities,
\bea
&&\frac{d\bar{P}_h}{dt}=-R(t)\bar{P}_h+R(t)\bar{P}_v,\nonumber\\
&&\frac{d\bar{P}_v}{dt}=R(t)\bar{P}_h-[R(t)+\Gamma(t)]\bar{P}_v.
\ea{11}
The rate that governs the polarization change in \r{11} is given by 
the integral
\be
R(t)=2\int_0^tdt'k(t,t')\theta(t,t'),
\e{12}
whose integrand is the product of 
\be
k(t,t')=\langle \mu(t)\mu(t')\rangle,
\e{52}
the correlation function of the random rotation rate $\mu(t)$, and of
\be
\theta(t,t')=\exp\left[-\frac{1}{2}\int_{t'}^t\Gamma(t_1)dt_1\right],
\e{10}
which is related to the polarizer transparency (measurement
effectiveness, as discussed below).
We assume that the average of $\mu(t)$ vanishes, 
$\overline{\mu(t)}$  $=$ $0$ (no systematic shift $\Delta\varphi$).

To obtain the validity condition of the above master equations, one
should consider higher-order terms in the cumulant expansion. 
As shown in Appendix \ref{A} for the case of continuous noise, the 
comparison of the second and fourth 
cumulants implies that Eqs. \r{11} hold under the condition
\be
R\ll\Gamma_{\rm R},
\e{22}
where $\Gamma_{\rm R}$ is the reciprocal correlation time of 
$\mu(t)$.
Below we assume that criterion \r{22} holds also for the case
of discrete noise.

\section{Continuous dephasing}
\label{Sec-continuous}

In this section we consider the case when $\bar{P}_{h,v}(t)$ {\em 
vary slowly on the time scale of several round trips},
which allows one to describe them by continuous functions of time on
the coarse-grained time scale
(a more general case will be discussed in Sec. \ref{Sec-discrete}).
In the present case the phase jumps 
\be
\Delta\varphi_n=\int_{(n-1)\tau_{\rm r}}^{n\tau_{\rm r}}dt\mu(t)
\e{19}
are small, $B^2$ $\equiv$ $\langle (\Delta\varphi_n)^2\rangle$  
$\ll$ $1$, and highly correlated, 
$\Delta\varphi_n$ $\approx$ $\Delta\varphi_{n-1}$, 
whereas the measurements are highly non-effective, $\theta\approx 1$. 
Then $\Gamma(t)$ can be substituted by 
\be
\Gamma_0=\frac{1}{\tau_{\rm r}}\int_0^{\tau_{\rm r}}dt\Gamma(t),
\e{13}
whereas $\mu(t)$ can be considered as a continuous, 
stationary random process, implying $k(t,t')=k(t-t')$. 
As a result, now in Eqs.~(\ref{11}) 
\be
R(t)=2\int_0^tdt'k(t')\exp\left(-\frac{\Gamma_0t'}{2} \right).
\e{14}

For $t\gg\Gamma_{\rm R}^{-1}$, where $\Gamma_{\rm R}$ is the
characteristic decay rate of $k(t)$, 
Eqs. \r{11} become Markovian (although they account
for non-Markovian phase fluctuations associated with $\mu(t)$)
\bea
&&\frac{d\bar{P}_h}{dt}=-R\bar{P}_h+R\bar{P}_v,\nonumber\\
&&\frac{d\bar{P}_v}{dt}=R\bar{P}_h-(R+\Gamma_0)\bar{P}_v,
\ea{15}
with {\em constant rate}
\be
R=2\int_0^\infty dtk(t)\exp \left(-\frac{\Gamma_0t}{2}\right) .
\e{16}
The last factor in the integrand here expresses the decay law 
of the first Stokes parameter, $f(t)=\bar{u}(t)/u(0)$, due to
measurement-induced relaxation.
This follows from Eqs. \r{6} [with $\mu(t)=0$] and \r{13}.

The solution of Eqs.~(\ref{15}) with the initial conditions 
\be
\bar{P}_h(0)=1,\qquad\bar{P}_v(0)=0,
\e{17}
yields
\be
\bar{P}_h(t)=e^{-(R+\Gamma_0/2)t}\left(\cosh St+\frac{\Gamma_0}{2S}
\sinh St\right),
\e{18}
where $S=\sqrt{R^2+\Gamma_0^2/4}$. 

In the absence of measurements, $\Gamma_0=0$, \r{18} yields
\be
\bar{P}_h(t)=\case{1}{2}(1+e^{-2R_0t}),
\e{20}
where $R_0=2\int_0^\infty dtk(t)$. 
By contrast, the measurement-affected polarization decays, assuming
that $\Gamma_0\gg R$, as
\be
\bar{P}_h(t)\approx e^{-Rt},
\e{21}
so that $R$ is indeed the measurement-modified decay rate.

The expression for the decay rate \r{16} can be recast in the same 
form as the universal result in Ref. \cite{Nature},
\be
R=2\pi\int_{-\infty}^\infty G(\omega)F(\omega)d\omega.
\e{23}
Here 
\be
G(\omega)=\frac{1}{\pi}\int_0^\infty dtk(t)\cos\omega t
\e{24}
is the random-field intensity spectrum, which can be considered as 
the spectrum of the {\em infinite-temperature reservoir}.
The other factor in \r{23},
\begin{eqnarray}
 F(\omega)=\frac{1}{\pi}\frac{\Gamma_0/2}{(\Gamma_0/2)^2+\omega^2},
\label{25}
\end{eqnarray}
is the Fourier transform of the measurement-induced decay law of the 
first Stokes parameter.
The width of $F(\omega)$ has the meaning of the effective rate of 
measurements $\nu$ \cite{Nature}.
The definition of $\nu$ introduced in 
\cite{Nature} becomes in the present case of a degenerate TLS
\be
\nu=[\pi F(0)]^{-1}.
\e{47}
From Eq. \r{25} one obtains $\nu=\Gamma_0/2$.

Equation (\ref{23}) allows a graphical interpretation
of the QZE\cite{Gil,kof99}. 
In particular, Eq. \r{23} shows that the {\em QZE occurs when the 
reservoir spectrum is peaked around} $\omega=0$.
However if {\em the spectral peak of the reservoir is detuned from 
the resonance frequency} $\omega_a$ of
the TLS (here $\omega_a=0$), one can obtain the quantum anti-Zeno
effect (AZE), i.e., an increase of the decay rate $R$ with the
effective measurement rate $\nu$, as illustrated in Sec. \ref{VA}.

\section{Discrete dephasing (phase jumps)}
\label{Sec-discrete}
\subsection{Small jumps: General analysis}
\label{VA}

Here we allow for any degree of correlation between consecutive phase 
jumps, as well as for arbitrary absorption per passage. 
In this subsection a general theory is developed for the case of
sufficiently small phase jumps.

We consider the sequence of polarization rotations by the angles 
$\Delta\varphi_n$ to be a discrete-time random process with the
correlation function $K_{nm}$ $=$ 
$\langle\Delta\varphi_n\Delta\varphi_m\rangle$. 
Typically, the random process $\Delta\varphi_n$ is {\em stationary}, 
yielding $K_{nm}=K_{n-m}=K_{m-n}$.
The general analysis is given in Appendix \ref{B}.
Here we present the simple case
\be
K_{n}=B^2\gamma^{|n|},
\e{26}
where $\gamma$ ($-1\le\gamma\le 1$) is the correlation degree between
two successive jumps\cite{kof90}: 
$\Delta\varphi_n\approx\Delta\varphi_{n+1}$ for $\gamma\approx 1$, 
$\Delta\varphi_n$ and $\Delta\varphi_{n+1}$ tend to have
opposite signs for $\gamma<0$ and are statistically
independent for $\gamma=0$. 
In this case [Eq. \r{26}] the correlation time is given by [cf. 
\r{27}]
\be
\Gamma_{\rm R}^{-1}=\frac{\tau_{\rm r}}{1-\gamma}.
\e{28}

We start from the master equations \r{11}, which are applicable both
for continuous and discrete evolution.
As shown in Appendix \ref{B}, under the conditions
\bem{37}
\bea
&&n\gg\frac{|\gamma|\theta}{1-(\gamma\theta)^2},\label{37a}\\
&&B^2\ll(1-\gamma)(1-\gamma\theta),
\ea{37b}
\eml
Eqs. \r{11} yield solution Eq. \r{18}, where now
\be
R=\frac{1+\gamma\theta}{1-\gamma\theta}\frac{B^2}{\tau_{\rm r}},
\qquad t=n\tau_{\rm r}.
\e{35}
For the general case
\be
R=\frac{1}{\tau_{\rm r}}\sum_{n=-\infty}^\infty K_n\theta^{|n|}.
\e{63}
One recognizes in this expression the analog of Eq. \r{16} for the 
discrete case, on realizing that $\theta^{n}$ ($n\ge 0$) is the 
discrete measurement-induced relaxation function of the first Stokes
coefficient.

\subsection{Correlated and anticorrelated discrete jumps: QZE and 
AZE}

\begin{figure}
\vspace{-2.5cm}
\centerline{\epsfig{file=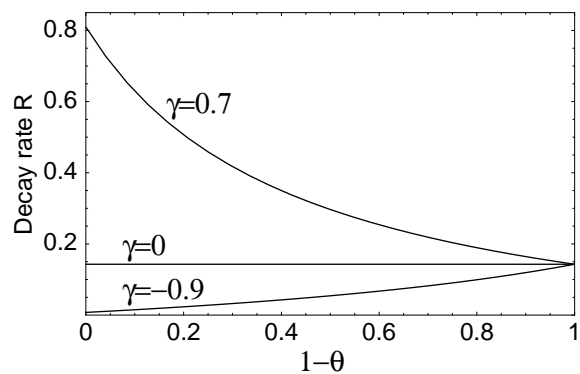,width=3.in}}
\vspace{-2.5cm}
\caption{Decay rate $R$ dependence on measurement effectiveness
$1-\theta$ for different degrees of correlation between consecutive
phase jumps, according to Eqs. \protect\r{35} and \protect\r{38}
Here $B=0.1,\ \tau_{\rm r}=0.07$. 
For the curves from top to bottom $\gamma=0.7$ (correlation leading to
QZE), 0, $-0.9$ (anticorrelation leading to AZE).
}\label{f2}\end{figure}

The decay rate $R$ \r{35} is independent of the measurements for
uncorrelated jumps, i.e., Markovian 
phase fluctuations ($\gamma=0$). However, $R$ as a function of
the effectiveness of measurements $1-\theta$ 
decreases for correlated phase jumps ($\gamma>0$), thus demonstrating 
the QZE, and increases for anticorrelated jumps ($\gamma<0$),
demonstrating the AZE (Fig. \ref{f2}).
How can we interpret these results?

The first step towards gaining insight into the decay rate \r{35} is
to realize that it can be recast in the form (Appendix \ref{B})
\be
R=2\pi\int_{-\pi/\tau_{\rm r}}^{\pi/\tau_{\rm r}}d\omega
G(\omega)F(\omega),
\e{38}
where
\be
G(\omega)=\frac{B^2}{2\pi\tau_{\rm r}}\frac{1-\gamma^2}{1+\gamma^2-
2\gamma\cos \omega\tau_{\rm r}}
\e{43}
is the reservoir spectrum, and
\be
F(\omega)=\frac{\tau_{\rm r}}{2\pi}\frac{1-\theta^2}{1+\theta^2-
2\theta\cos \omega\tau_{\rm r}}
\e{40}
is the Fourier transform of the measurement-induced relaxation
function of the first Stokes parameter [see Eq. \r{B1}].
The form \r{38} is essentially the same as \r{23} or the universal
form of measurement-affected decay rates given in Ref. \cite{Nature}.
What makes \r{38} distinct is that we are now in a finite frequency 
domain, $-\pi/\tau_{\rm r}<\omega<\pi/\tau_{\rm r}$, in contrast to
Eq. \r{23}. This is related to the fact that polarization evolution
studied here is discrete in time. 
In what follows, we shall first separately analyze $F(\omega)$ and
$G(\omega)$, and then graphically deduce the QZE and AZE from their
convolution \r{38}.

\subsubsection{Properties of $F(\omega)$}

The Fourier-transformed measurement-induced relaxation function of the
first Stokes parameter \r{40} is normalized to one. 
It has one maximum,
$F_{\rm max}(\omega=0)=(\tau_{\rm r}/2\pi)(1+\theta)/(1-\theta)$
located at $\omega=0$.
It is minimal at the borders $\omega=\pm\pi/\tau_{\rm r}$ of the
frequency domain, 
$F_{\rm min}=(\tau_{\rm r}/2\pi)(1-\theta)/(1+\theta)$. 
In the particular case of the ideal (fully effective) projective 
measurements (IMs), $F(\omega)$ is constant,
\be
F(\omega)=\frac{\tau_{\rm r}}{2\pi}\qquad(\theta=0),
\e{45}
whereas for {\em non-effective} (unreliable) measurements 
($\theta\approx 1$), $F(\omega)$ is a narrow peak [cf. Eq. \r{25}],
\bea
F(\omega)\approx\frac{1}{\pi}\frac{\Gamma_0/2}
{(\Gamma_0/2)^2+\omega^2}\qquad(|\omega|\tau_{\rm r}\ll 1).
\ea{44}
Here we took into account that now 
$(1-\theta)\approx\Gamma_0\tau_{\rm r}/2\ll 1$ [cf. Eqs. \r{3} and 
\r{13}].

Inserting Eq. \r{40} into \r{47}, one obtains that the effective rate
of measurements is
\be
\nu=\frac{2(1-\theta)}{1+\theta}\frac{1}{\tau_{\rm r}}.
\e{48}
For any allowed value of $\theta$, we have 
$\nu\sim(1-\theta)/\tau_{\rm r}$,
which formally confirms the intuitive interpretation of
$1-\theta$ as the effectiveness of measurements.
More specifically, for IMs, $\nu=2/\tau_{\rm r}$, which differs by a
factor of 2 from the real rate of measurements (note that the
definition of $\nu$ is meaningful only with an accuracy up to a factor
of the order of one). 
For low-effectiveness (highly unreliable) measurements 
($\theta\approx 1$), we have $\nu=(1-\theta)/\tau_{\rm r}=\Gamma_0/2$.

\subsubsection{Properties of $G(\omega)$}

The reservoir-coupling spectrum \r{43} is constant
in the Markovian case $G(\omega)$, $\gamma=0$,
\be
G(\omega)=\frac{B^2}{2\pi\tau_{\rm r}}\qquad(\gamma=0).
\e{41}
By contrast, $G(\omega)$ is mainly concentrated near $\omega=0$ for 
{\em highly correlated} jumps ($\gamma\approx 1$),
\be
G(\omega)\approx\frac{B^2}{\pi\tau_{\rm r}^2}\frac{\Gamma_{\rm R}}
{\Gamma_{\rm R}^2+\omega^2}\qquad(|\omega|\tau_{\rm r}\ll 1),
\e{46}
whereas for highly anticorrelated jumps ($\gamma\approx -1$), it is 
peaked near $\omega=\pm\pi/\tau_{\rm r}$,
\bea
G(\omega)\approx\frac{B^2}{\pi\tau_{\rm r}^2}\frac{\Gamma_{\rm R}'}{
{\Gamma_{\rm R}'}^2+(\pi/\tau_{\rm r}\pm\omega)^2}\quad
(\pi\pm\omega\tau_{\rm r}\ll 1),
\ea{42}
with $\Gamma_{\rm R}'=(1+\gamma)/\tau_{\rm r}$. 

\subsubsection{Graphical analysis of the decay rate}

\begin{figure}
\vspace*{-3.cm}
\centerline{\epsfig{file=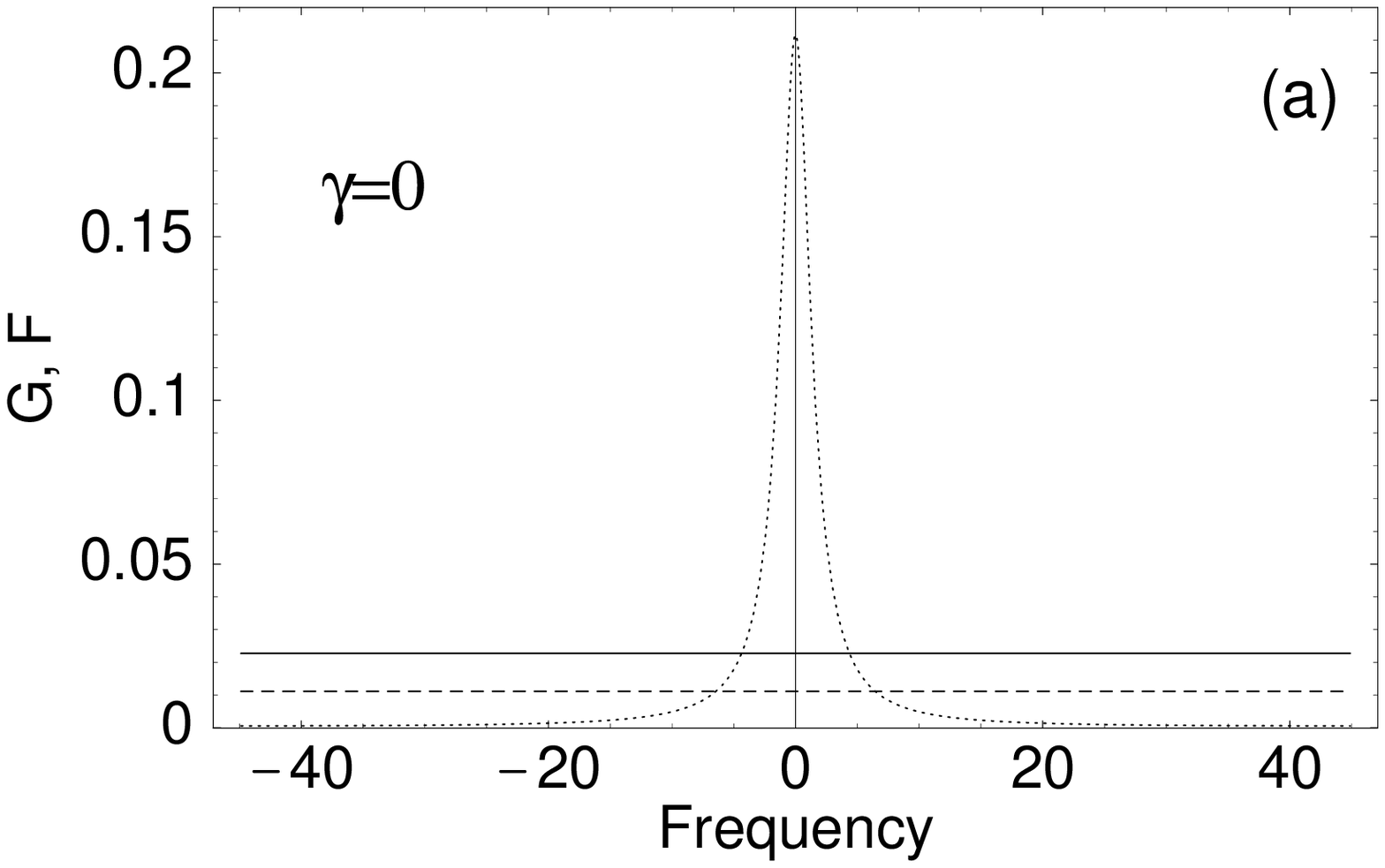,width=3.in}}
\vspace*{-5cm}
\centerline{\epsfig{file=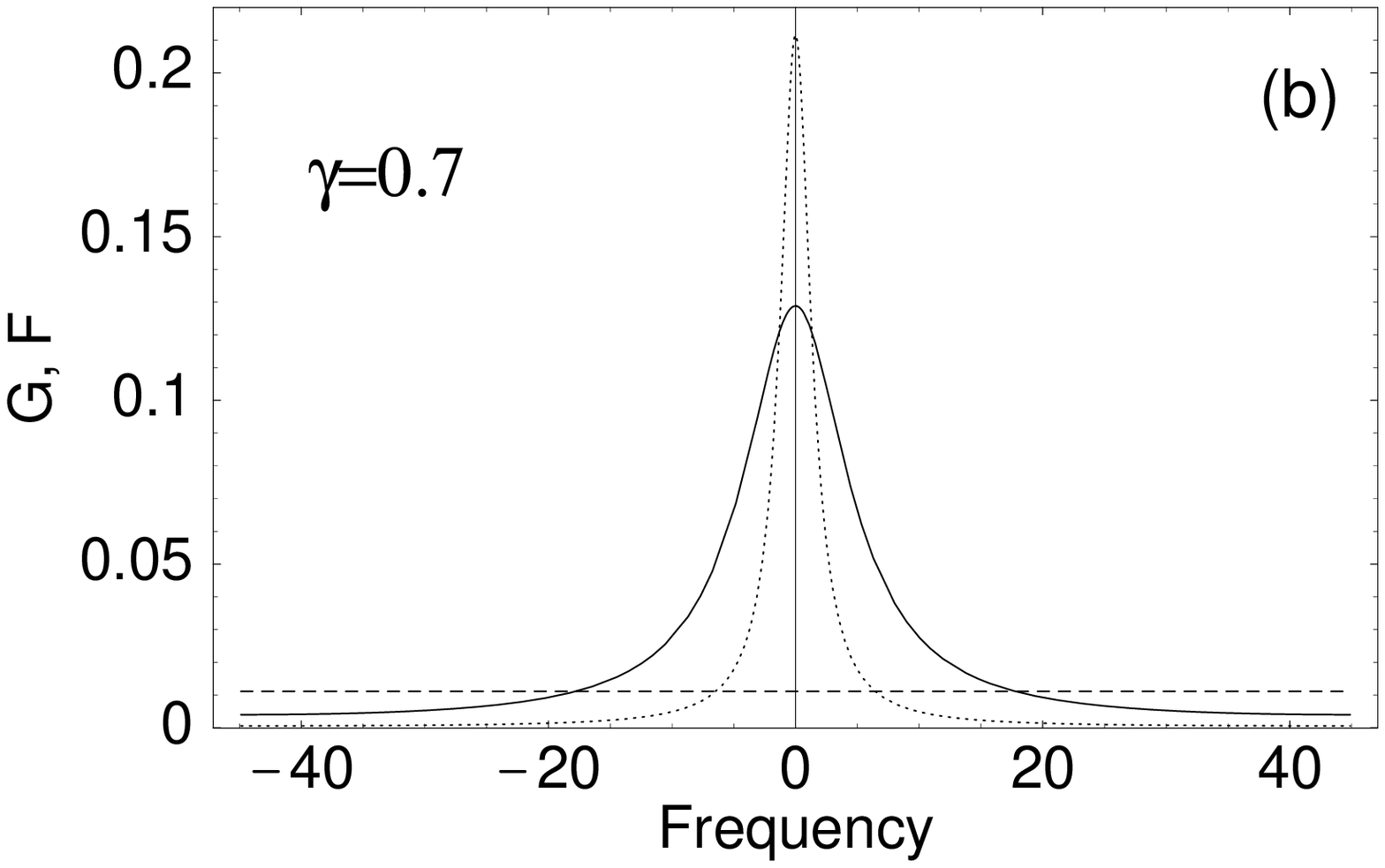,width=3.in}}
\vspace*{-5cm}
\centerline{\epsfig{file=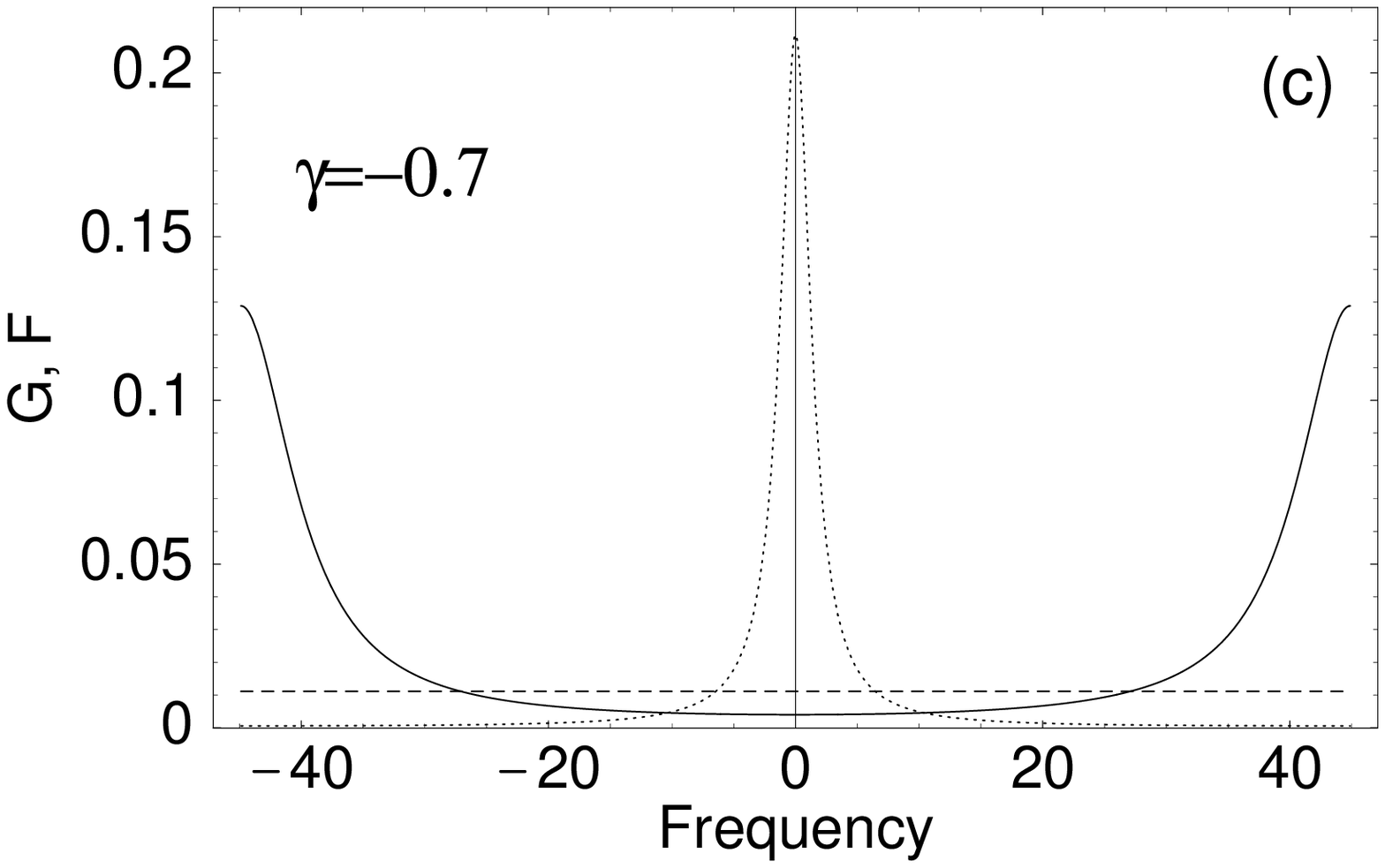,width=3.in}}
\vspace{-2.5cm}
\caption{Overlap of the reservoir spectrum $G(\omega)$ in 
\protect\r{43} and the repeated-measurement broadening $F(\omega)$ in 
\protect\r{40}, determining the convolution integral of the decay 
rate $R$ in \protect\r{32}.
Solid lines: $G(\omega)$ with (a) $\gamma=0$ (uncorrelated jumps), 
(b) $\gamma=0.7$ (correlated jumps - QZE),
(c) $\gamma=-0.7$ (anticorrelated jumps - AZE). 
Dashed lines: $F(\omega)$ with $\theta=0$ (perfect projections). 
Dotted lines: $F(\omega)$ with $\theta=0.9$ (ineffective 
measurements). 
Here $B=0.1,\ \tau_{\rm r}=0.07$. 
}\label{f3}
\end{figure}

Since the decay rate  Eq. \r{38} is determined by the overlap of
$F(\omega)$ and $G(\omega)$, the above results allow a graphical
interpretation of the dependence of $R$ on the effective measurement
rate (Fig. \ref{f3}).
In Fig. \ref{f3}(a), $G(\omega)$ is flat ($\gamma=0$) and the 
convolution \r{38} is proportional to the integral of $F(\omega)$,
$\int_{-\pi/\tau_{\rm r}}^{\pi/\tau_{\rm r}}d\omega F(\omega)=1$.
As a result, $R$ is independent of the shape of $F(\omega)$ (i.e., of 
the effectiveness of measurements).
In Fig. \ref{f3}(b) ($\gamma>0$), $G(\omega)$ is peaked at $\omega=0$
and hence the convolution $R$ is determined by the portion of 
$F(\omega)$ inside the width of $G(\omega)$.
This implies a reduction of $R$ with the broadening of $F(\omega)$,
i.e., the QZE.
The opposite is true in Fig. \ref{f3}(c),
due to the fact that the reservoir
peaks are detuned from the TLS frequency $\omega_a=0$.
Now the convolution $R$ is sensitive to the wings of $F(\omega)$,
which rise with the increase of $F(\omega)$.
As a result, the AZE occurs (increase of $R$ with the broadening of 
$F(\omega)$, i.e., with the increase of the measurement 
effectiveness).
In Figs. \ref{f3}(b), \ref{f3}(c), the limit of the flat $F(\omega)$ 
(i.e., IMs) implies that the integral \r{38} is independent of the 
shape of $G(\omega)$, resulting in the same value of $R$ as in the 
Markovian case, Fig. \ref{f3}(a).

The decay rate $R$ corresponding to Eq. \r{5} can be recast as 
\be
R=2(\df/\tau_{\rm r})^2/\nu,
\e{51} 
with the effective rate of measurements $\nu$ given by Eq. \r{48}.
The above expression for $R$ has the familiar form characteristic of 
the QZE \cite{Nature}.
The coefficient $(\df/\tau_{\rm r})^2=\bar{\mu}^2$ is the squared
average of the coupling amplitude over the round trip.

\section{Exactly solvable models of random phase jumps}
\label{VI}

We now consider simple models allowing an exact solution in the
cases of {\em absent measurements (free evolution)\/} and 
{\em perfect measurements\/} (IMs)].

In the case of free evolution the probability that the photon is found
with the horizontal polarization is given by 
\be
P_h(n)=\langle\cos^2\varphi_n\rangle=\frac{1}{2}+\frac{1}{2}\text{Re}
\langle e^{2i\varphi_n}\rangle,
\e{int1}
where $\varphi_n=\sum_{k=1}^n\Delta\varphi_k$ and the angle brackets
denote the averaging.

\subsection{Markovian limit (uncorrelated jumps)}

First, let us assume that the rotation angles $\Delta\varphi$ during 
the round-trips are independent of each other. 
We can then write Eq. \r{int1} in the form
\begin{eqnarray}
  P_{h}(n)& =& \frac{1}{2} + \frac{1}{2}\text{Re}\int 
  p(\Delta\varphi_1) \dots \int p(\Delta\varphi_{n})\nonumber\\ 
  &\times&\exp\left[2i\sum_{k=1}^{n}
  \Delta\varphi_{k}\right] d\Delta\varphi_1 \dots d\Delta\varphi_n.
\end{eqnarray}
Assuming the probability distribution of each $\Delta\varphi_k$ to be
symmetric, we finally arrive at
\begin{eqnarray}
  P_{h}(n) = \frac{1}{2} + \frac{1}{2} \langle \cos 2 \Delta\varphi_k 
  \rangle ^n ,
  \label{prob2}
\end{eqnarray}
with
\begin{eqnarray}
  \langle \cos 2 \Delta\varphi_k \rangle  = \int p(\Delta\varphi_k) 
  \cos 2 \Delta\varphi_kd\Delta\varphi_k .
  \label{mean}
\end{eqnarray}
If, for simplicity, we assume that the rotation angle can take just 
two values $\pm \Delta \varphi$, each with the probability
1/2, then Eq. (\ref{prob2}) reduces to
\begin{eqnarray}
  P_{h}(n) = \frac{1}{2} + \frac{1}{2}  \cos ^n 2 \Delta \varphi  .
  \label{prob3}
\end{eqnarray}
Equations (\ref{prob2}) and (\ref{prob3}) describe a purely 
exponential Markovian decay,
for which we cannot expect any inhibition by the QZE.

\subsection{Simple non-Markovian model}
\subsubsection{Description of the model}
\label{VIB1}

Let us now consider a simple model for discrete non-Markovian 
dephasing, in which the probability of the rotation angle in the 
$n$th step depends on the rotation angle in the $(n-1)$th step (the 
so-called ``random walk with persistence'' \cite{kam81}). 

We assume that the rotation step can take two values 
$\pm \Delta \varphi$ with equal probabilities
$p_0(\pm \Delta \varphi)=1/2$, whereas the $n$th rotation angle is 
equal or opposite to the previous one $(n-1)$ with the probability 
$p$ or $q=1-p$, respectively.
Correspondingly, the conditional probabilities 
$P(\Delta\varphi_{n+1},\Delta\varphi _{n})$ are defined by
\begin{eqnarray}
  \label{conpr1}
  P(\Delta\varphi _{n},\Delta\varphi _{n}) = p , \qquad
  P(-\Delta\varphi _{n},\Delta\varphi _{n}) &=& q.
  \label{conpr2}
\end{eqnarray}
As a consequence, for a given $\Delta\varphi_{n}$ the conditional 
average of $\Delta\varphi_{n+1}$ is
\bem{49}
\bea
&&\langle\Delta\varphi_{n+1}\rangle_{\Delta\varphi_{n}}=
\gamma\Delta\varphi_{n},\label{49a}\\
&&\gamma=2p-1.
\ea{49b}
\eml
The rotation steps are correlated for $1/2<p\le 1$
($\gamma>0$), anticorrelated for $0\le p<1/2$ ($\gamma<0$), and
uncorrelated for $p=1/2$ ($\gamma=0$).
Equation \r{49a} implies that the correlation function is given by 
Eq. \r{26} with
\be
B=\df.
\e{50}

For small jumps and large $n$, Eq. \r{37}, the evolution of the 
probability of horizontal polarization can be approximated by 
combining Eqs. \r{18}, \r{35}, and \r{50}.

\subsubsection{Free evolution}
In the case of free evolution, Eqs. \r{57} and \r{58} in Appendix 
\ref{C} yield the analytical solution 
\be
P_h(n)=\frac{1}{2}+\frac{g(r)-g(-r)}{4r},
\e{nonmarc1}
where $g(r)=(q\cos 2\Delta\varphi+r)(p\cos 2\Delta\varphi+r)^n$ and
$r=\sqrt{q^2-p^2\sin^22\Delta\varphi}$.

We now consider special cases of Eq. \r{nonmarc1}.
If $p$ $=$ $q$ $=$ $1/2$ we get the Markovian
exponential decay given by Eq. \r{prob3}. 
For $4\Delta\varphi^2\ll 1,q^2/p^2$ one obtains 
$r\approx q-2p^2\Delta\varphi^2/q\approx q$ and Eq. \r{nonmarc1}
reduces to Eq. \r{20} with $R_0=p(\df)^2/q\tau_{\rm r}$, which, in
view of Eqs. \r{49b} and \r{50},
reduces to the formula for $R$ in \r{35} with $\theta=1$.
The corresponding behavior can be interpreted as a random sequence of 
``independent'' rotations of the length $\Delta \varphi p/q$, each 
rotation taking place after $p/q$ steps. 
Using the same arguments as for the derivation of Eq. (\ref{prob3}), 
we arrive at 
\begin{eqnarray}
  P_{h}(n) \approx \frac{1}{2} + \frac{1}{2} \left[ \cos  \left( 2 
  \Delta \varphi 
  \frac{p}{q} \right) \right] ^{\frac{qn}{p}}  .
  \label{prob4}
\end{eqnarray}
Finally, for $p$ $=$ 1 (i.e., $\gamma=1$) the rotation steps are 
infinitely long -- the exponential approximation is never true and 
Eq. \r{nonmarc1} yields the ``Rabi oscillations'' \r{4}. 


\subsubsection{``Impulsive'' (projective) measurements.}
Let us now assume that during each round-trip  the polarization 
state of the photon is projected on the horizontal polarization 
($\theta=0$ in \r{3}). 
The total probability of the horizontal polarization is  given by 
Eq. (\ref{expdecay}), independently of the memory of the rotator.

Since for $p$ $>$ $q$ this exponential decay is slower than the 
non-Markovian decay (\ref{prob4}) [or, more exactly,
(\ref{nonmarc1})], there exists a region of $n$ values for 
which the probability of the horizontal polarization is larger with 
the projection than without it. 
(The region is finite because (\ref{prob4}) and (\ref{nonmarc1}) 
decay towards 1/2 whereas (\ref{expdecay}) decays to zero.)

\begin{figure}[htb]
\centerline{\epsfig{file=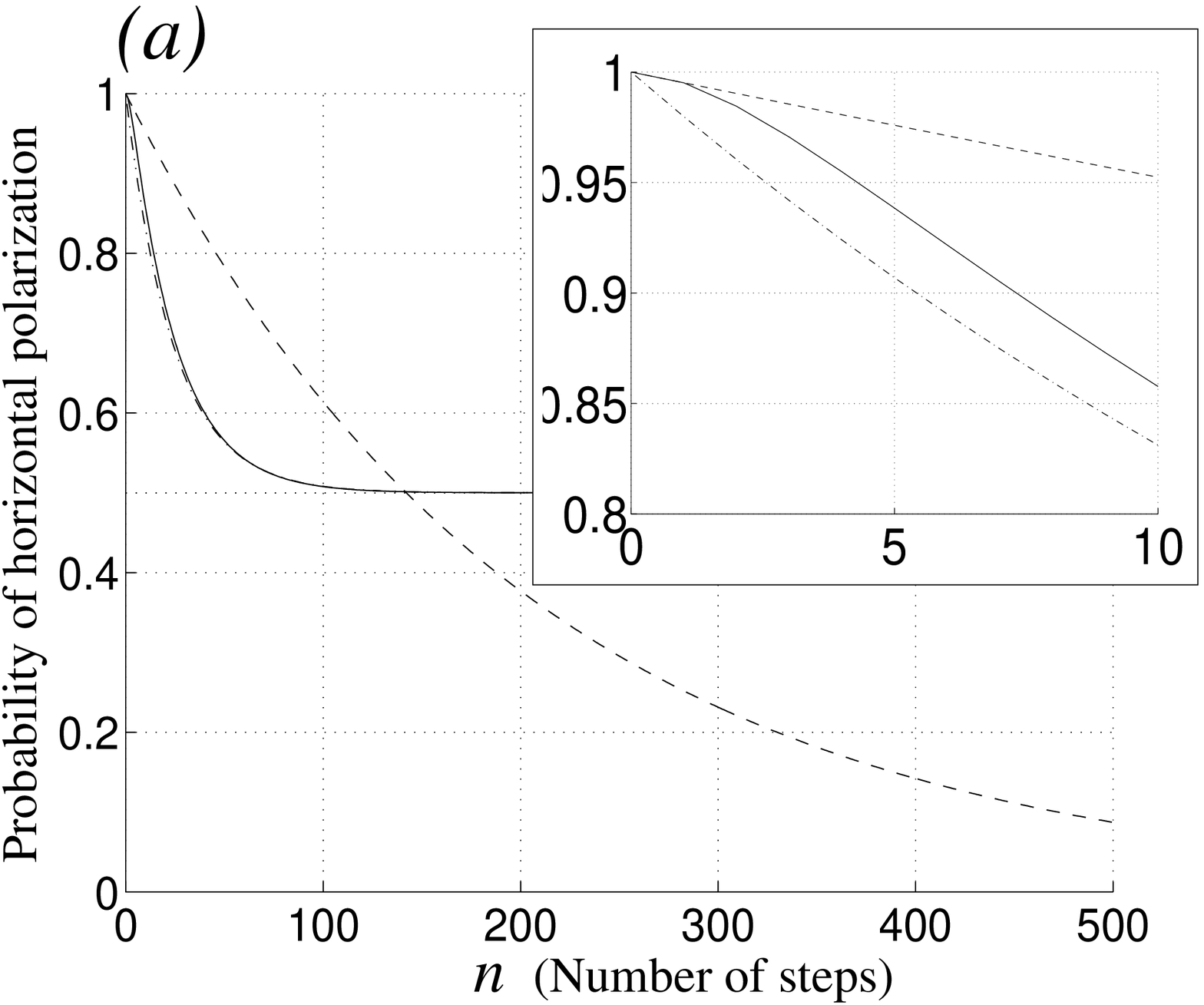,width=2.5in}}
\centerline{\epsfig{file=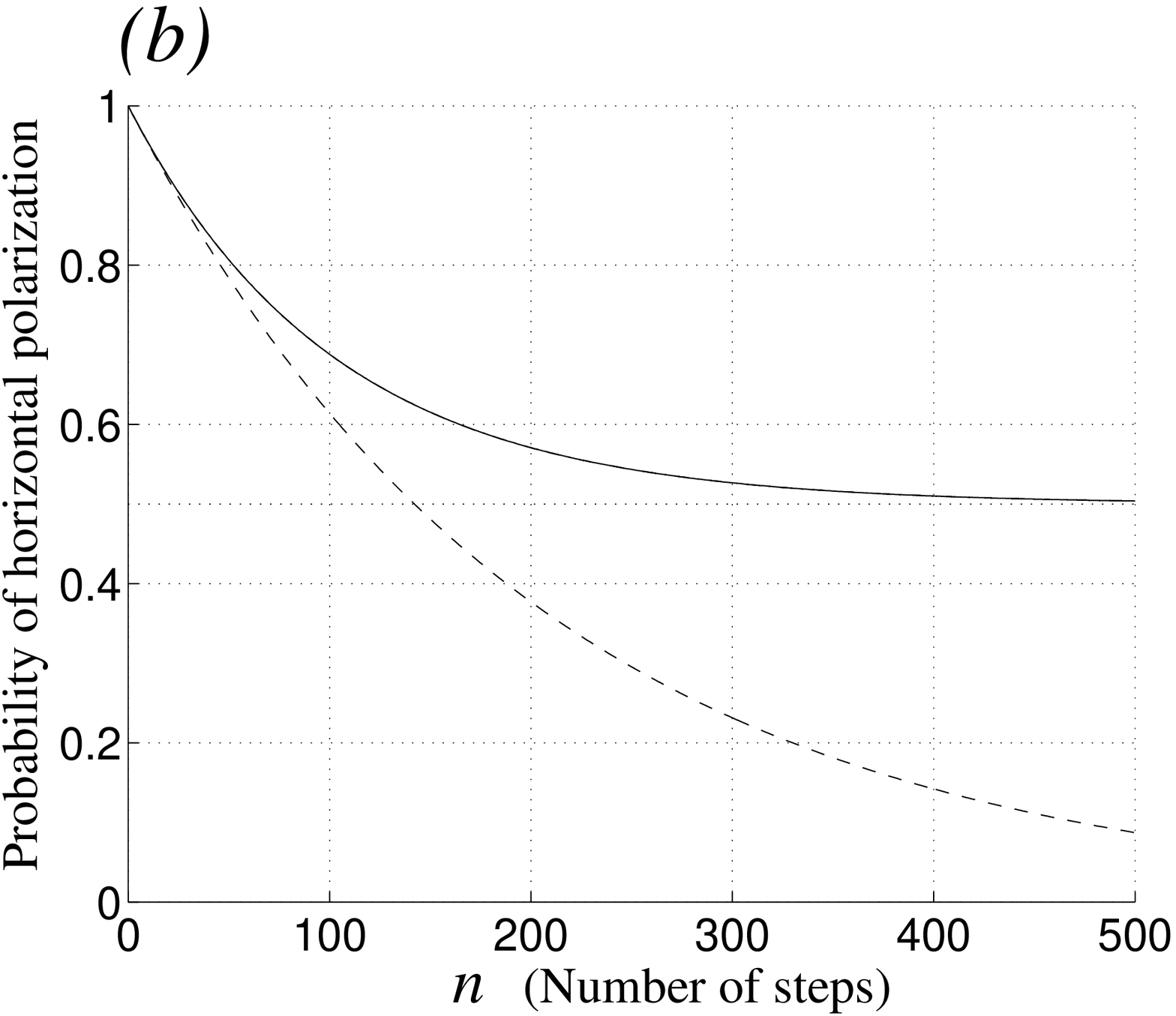,width=2.5in}}
\centerline{\epsfig{file=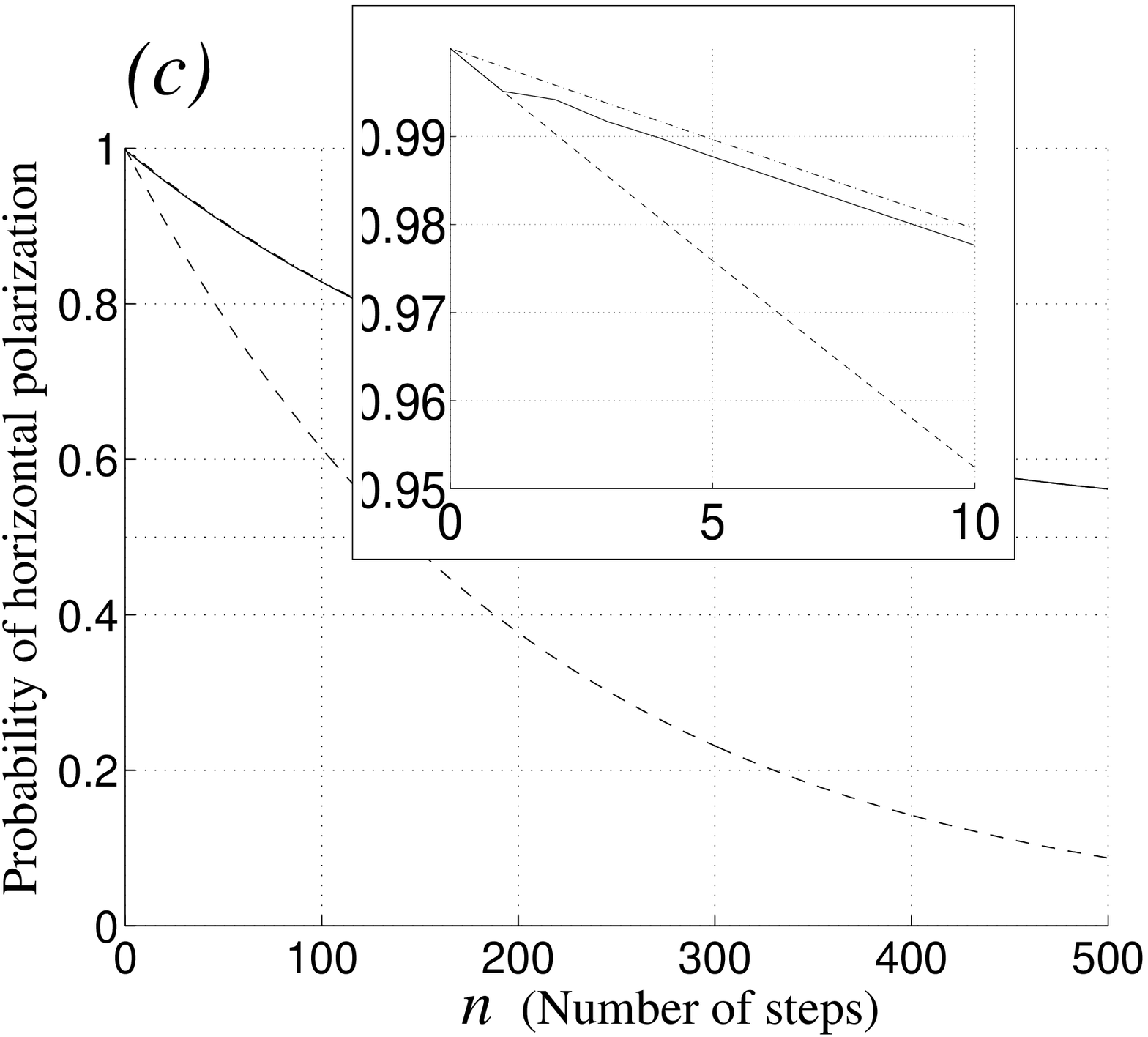,width=2.5in}}
\vspace{.5cm}
\protect\caption{Evolution of the probability of horizontal 
polarization for $\df=4^\circ$: (a) $p=0.8$, (b) $p=0.5$, (c) $p=0.3$,
as a function of the number of round trips $n$.
Inset (a): same parameters as (a) but for smaller $n$.
Inset (c): same parameters as (c) but for smaller $n$.
Solid line, the probability of the horizontal 
polarization without measurements, Eq. (\protect\ref{nonmarc1});
dash-dotted line, the exponential approximation
(\protect\ref{prob4}); dashed line, decay with perfect measurements 
(complete absorption), as in Eq. (\protect\ref{expdecay}). 
}
\label{fig2}
\end{figure}

Figures \ref{fig2}(a)-(c) represent the discussed model.
Figure \ref{fig2}(a) exhibits the QZE.
In Fig.  \ref{fig2}(b) we see the case of exponential Markovian 
decay ($p$ = 1/2) for which the QZE does not take place.
It differs from the uninterrupted evolution, which tends to 
equilibrium of the population. 
Figure \ref{fig2}(c) exhibits the AZE for $p<1/2$.

Generally, the smaller $\Delta \varphi$ the slower are thermalization 
and decay.
For $1/2<p<1$ (corresponding to $0<\gamma<1$ in \r{49b}; cf. Sec. 
\ref{VA}) the larger $p$ the larger is decay inhibition by the 
measurements (QZE), whereas for $0<p<1/2$ ($-1<\gamma<0$) the 
opposite trend, i.e., decay acceleration (AZE) occurs.
The generic properties of the QZE and the AZE clearly follow from 
this very simple model. 

\section{Conclusions}
\label{Sec-concl}

We have obtained the general conditions of decay inhibition or 
acceleration via polarization measurements. 
This has been accomplished by expressing the decay rate as a 
convolution of two functions: (i) the fluctuation spectrum of the
random polarization-rotation rate, which is analogous to the spectrum 
of the infinite-temperature reservoir; (ii) the Fourier transform of 
the measurement-induced polarization dephasing, which is determined 
by the absorber transparency. 
Inhibition of the decay by frequent measurements (the QZE) has been
shown to occur when the reservoir spectrum is peaked around zero 
frequency, which corresponds to correlated polarization-angle jumps. 
However, if the jumps are anti-correlated, the reservoir spectrum is 
split into two peaks, with a dip at zero frequency. 
Then one should obtain the quantum AZE, i.e., an increase of the 
decay rate as the rate of effective measurements increases.

In \cite{Nature} only {\em direct} measurement schemes were 
considered (Cook's \cite{coo88} scheme with pulsed or cw measuring 
field), whereas here we have considered an {\em indirect destructive} 
measurement scheme of the polarization state.
Our analysis confirms that the general formula in \cite{Nature} holds 
for {\em direct and indirect} measurement schemes, irrespective of 
their differences, 
provided they approximately yield projections on the measured state.
The general formula in \cite{Nature} for the effect of frequent 
measurements on decay was claimed to hold not just for ideal 
measurements, i.e., instantaneous projections, but also for 
continuous measurements. 
Generally, measurements can be ineffective (i.e., 
producing projections with a probability less than one) \cite{per90}.
Here we have considered a comprehensive measurement model, which 
encompasses the various types of measurements: ideal, ineffective 
(impulsive), and continuous.
The present analysis corroborates the generality of the formula in 
\cite{Nature} and provides unifying expressions for the 
measurement-induced broadening function $F(\omega)$ [Eq. \r{40}] and 
the effective measurement rate $\nu$ [Eq. \r{48}].

The significance of the present analysis lies in the physical 
simplicity of the model and its experimental realizability, as well 
as in the ability to control the reservoir spectrum (or memory time) 
and measurement rate by adjustment of the Pockels-cell modulation (on
a ns scale) and the PBS transparency, respectively. 
It allows us to test theoretically (and, hopefully, experimentally 
in the future) the general conclusions from Ref.~\cite{Nature} for the
Zeno and anti-Zeno effects.

\acknowledgments
This work was supported by ISF (G. K.), 
the Ministry of Absorption (A. K.) and Minerva (T. O. and G. K.).

\appendix
\section{Derivation of the master equations and their validity
conditions}
\label{A}

To derive the equation for the average solution of Eq. \r{6} we use
the cumulant expansion technique \cite{ter74,kof90}.
Assuming, for simplicity, that the odd moments of $\mu(t)$ vanish, we
obtain
\be
\frac{d\bar{Q}}{dt}=A(t)\bar{Q}+\int_0^tdt'K(t,t')\bar{Q}(t'),
\e{A1}
where 
\be
K(t,t')=\Theta_2(t,t')+
\int_{t'}^tdt_1\int_{t'}^{t_1}dt_2\Theta_4(t,t_1,t_2,t')+\dots.
\e{A2}
Here the totally ordered cumulants are
\bem{A3}
\bea
&\Theta_2&(t,t')=k(t,t')CU(t,t')C,\label{A3a}\\
&\Theta_4&(t,t_1,t_2,t')=M_4(t,t_1,t_2,t')CU(t,t_1)CU(t_1,t_2)
\nonumber\\
&&\times CU(t_2,t')C
\ea{A3b}
\eml
where $U(t,t')=\exp[\int_{t'}^td\tau A(\tau)]$ and 
\be
M_4(t,t_1,t_2,t')=\langle\mu(t)\mu(t_1)\mu(t_2)\mu(t')\rangle
-k(t,t_1)k(t_2,t')
\e{A7}
is the the fourth cumulant of $\mu(t)$,
the angular brackets denoting the average.

Truncating the cumulant expansion \r{A2} at the second order, one
obtains the following integro-differential equations for the elements
of the polarization tensor,
\bem{A4}
\bea
&\frac{d\bar{P}_h}{dt}=&-2\int_0^tdt'k(t,t')\theta(t,t')
[\bar{P}_h(t')-\bar{P}_v(t')],\label{A4a}\\
&\frac{d\bar{P}_v}{dt}=&2\int_0^tdt'k(t,t')\theta(t,t')[\bar{P}_h(t')
-\bar{P}_v(t')]\nonumber\\
&&-\Gamma(t)\bar{P}_v,\label{A4b}\\
&\frac{d\bar{u}}{dt}=&-\frac{\Gamma(t)}{2}\bar{u}
-2\int_0^tdt'k(t,t')[1+\theta^2(t,t')]\bar{u}(t').
\ea{A4c}
\eml
Note that Eqs. \r{A4a} and \r{A4b} for the polarization probabilities 
are uncoupled from Eq. \r{A4c} for the coherence.

We assume that $\bar{P}_h(t')$ and $\bar{P}_v(t')$ are slowly 
changing with respect to the integral kernel. 
Then one can set $\bar{P}_{h,v}(t')\approx\bar{P}_{h,v}(t)$ in Eqs.
\r{A4a} and \r{A4b}, yielding Eqs. \r{11}.

Consider now the validity conditions of Eqs. \r{11} for the case when
$\mu(t)$ is a continuous-time stationary random process and
$\Gamma(t)=\Gamma_0$ is constant.
First, the condition to transform Eqs. \r{A4a} and \r{A4b} into Eqs. 
\r{11} can be seen to be 
\be
R\ll\Gamma_{\rm R}+\Gamma_0.
\e{A5}
Second, one can show that taking into account the fourth cumulant
\r{A3b} in Eqs. \r{A1} and \r{A2} amounts to the addition of the term
\bea
&-4\int_{t'}^tdt_1\int_{t'}^{t_1}dt_2&M_4(t,t_1,t_2,t')
\theta(t,t_1)\theta(t_2,t')\nonumber\\
&&\times[1+\theta^2(t_1,t_2)]
\ea{A6}
to the kernel $2k(t,t')\theta(t,t')$ of Eqs. \r{A4a} and \r{A4b}.
The cumulant $M_4(t,t_1,t_2,t')$ tends to zero with the increase 
of the difference between any two of its arguments.
Assuming that all the characteristic decay rates are of the order of 
$\Gamma_{\rm R}$, one can estimate that the fourth cumulant
contributes to the rate $R$ in Eqs. \r{15} a quantity of the order of
$R^2/\Gamma_{\rm R}$.
Hence, the fourth cumulant can be neglected under condition \r{22}.
Note that condition \r{22} is stricter than \r{A5}.

\section{General analysis of small discrete phase jumps}
\label{B}

The correlation time $\Gamma_{\rm R}^{-1}$ of a random chain
$\df_n$ can be defined generally as
\be
\Gamma_{\rm R}^{-1}=\frac{\tau_{\rm r}}{B^2}\sum_{n=0}^{\infty}K_n.
\e{27}

Assuming the validity of condition \r{22},
the solution of Eqs. \r{11} with the initial conditions \r{17} 
can be obtained at $t=n\tau_{\rm r}$ in the form
\be
\left(\begin{array}{c}\bar{P}_h(t)\\ 
\bar{P}_v(t)\end{array}\right)=\exp\left[
\left(\begin{array}{rr}-W_n&W_n\\
W_n&\ -W_n-\Gamma_0t\end{array}\right)\right]
\left(\begin{array}{c}1\\ 
0\end{array}\right),
\e{29}
where $W_n=\int_0^tdt'R(t')$.

To prove Eq. \r{29} we cast Eqs. \r{11} in the form $\dot{q}=L(t)q$,
where
\be
q(t)=\left(\begin{array}{c}\bar{P}_h(t)\\
\bar{P}_v(t)\end{array}\right),\qquad
L(t)=\left(\begin{array}{lr}-R(t)&R(t)\\
R(t)&-R(t)-\Gamma(t)\end{array}\right).
\e{60}
Hence $q(t)=U(t)q(0)$, where $U(t)=T\exp[\int_0^tdt'L(t')]$, $T$ being
the time-ordering operator.
Note that $L(t)\approx L(\infty)=L$ for $t\gg\Gamma_{\rm R}^{-1}$.
Consider two cases.
Let, first, $\Gamma_0\ll\Gamma_{\rm R}$.
We want to show that $T$ can be omitted in the expression for $U(t)$,
resulting in the coincidence of $U(t)$ with the matrix in Eq. \r{29}.
For $t\ll(R+\Gamma_0)^{-1}$ one gets 
$U(t)\approx[1+\int_0^\tau dt'L(t')]\approx\exp[\int_0^tdt'L(t')]$,
whereas for $t\agt(R+\Gamma_0)^{-1}$, choosing time $\tau$ such that 
$\Gamma_{\rm R}^{-1}\ll\tau\ll(R+\Gamma_0)^{-1}$,
one obtains $U(t)\approx[1+\int_0^\tau dt'L(t')]e^{L(t-\tau)}\approx
\exp[\int_0^tdt'L(t')]$.
The second case $\Gamma_0\agt\Gamma_{\rm R}$ implies, in view of
\r{22}, that $\Gamma_0\gg R$.
Then one can set $\bar{P}_v(t)\approx 0$ in Eqs. \r{11}, yielding
$\bar{P}_h(t)\approx e^{-W_n}$, which can be shown to follow also from
\r{29}. 
This finishes the proof of \r{29}.

The above expression for $W_n$, with the account of Eqs. \r{12}-\r{10}
and \r{3}, can be cast as
\be
W_n=\sum_{m,m'=1}^nK_{mm'}\theta^{|m-m'|}
\e{31}
or, for a stationary process,
\be
W_n=nB^2+2\sum_{m=1}^{n-1}\sum_{m'=1}^mK_{m'}\theta^{m'}.
\e{32}

In the case \r{26}, the calculation of \r{32} shows that, under 
condition \r{37a},
\be
W_n\approx nB^2\frac{1+\gamma\theta}{1-\gamma\theta}-\frac{2B^2
\gamma\theta}{(1-\gamma\theta)^2}.
\e{33}
For
\be
B^2\ll(1-\gamma\theta)^2
\e{34}
the second term in the right-hand side of Eq. \r{33} can be neglected
and Eq. \r{29} yields \r{18} with the definitions \r{35}. 

For an arbitrary correlation function $K_n$ one can show with the help
of Eq. \r{32} that Eq. \r{35} for $R$ should be substituted by
\be
R=\frac{1}{\tau_{\rm r}}\left(B^2+2\sum_{n=1}^\infty 
K_n\theta^n\right)
\e{36}
or, equivalently, by Eq. \r{63}.

The validity conditions of the above results are: 
$t\gg\Gamma_{\rm R}^{-1}$ [roughly corresponding to inequality  
\r{37a}], the inequalities \r{22} and $B^2\ll 1$. 
In the case \r{26} the latter two inequalities can be combined to 
obtain Eq. \r{37b}, which is generally stricter than \r{34}.

Equation \r{63} can be rewritten in the form \r{38} with
\be
F(\omega)=\frac{\tau_{\rm r}}{2\pi}\sum_{n=-\infty}^\infty\theta^{|n|}
e^{-in\omega\tau_{\rm r}}
\e{B1}
and
\bea
G(\omega)&=&\frac{1}{2\pi\tau_{\rm r}}\sum_{n=-\infty}^\infty K_n
e^{in\omega\tau_{\rm r}}\nonumber\\
&=&\frac{1}{2\pi\tau_{\rm r}}\left(B^2+
4\sum_{n=1}^\infty K_n\cos n\omega\tau_{\rm r}\right).
\ea{39}
The function $F(\omega)$ \r{B1} is normalized to one in the interval
$-\pi/\tau_{\rm r}<\omega<\pi/\tau_{\rm r}$. 
Performing the summation in Eq. \r{B1} yields Eq. \r{40}.
In the special case \r{26} Eq. \r{39} reduces to the closed formula
\r{43}.

\section{Exact solution for discrete non-Markovian phase jumps}
\label{C}

Here we calculate polarization dephasing for a rather general model of
random rotation angles.
The average phase factor in Eq. \r{int1} is given generally by
\be
\langle e^{2i\varphi_n}\rangle=
\sum_{\Delta\varphi_1,\dots,\Delta\varphi_n}e^{2i\varphi_n}
p(\Delta\varphi_1,\dots,\Delta\varphi_n),
\e{53}
where $p(\Delta\varphi_1,\dots,\Delta\varphi_n)$ is the joint
probability of the random chain $\Delta\varphi_k$.

According to the model of Sec.\ref{VIB1}, 
$\varphi_n=\sum_{k=1}^n\Delta\varphi_k$ is a (non-Markovian) sum over
a Markovian chain $\Delta\varphi_k$.
The average \r{53} can be calculated for arbitrary Markovian chains,
i.e., chains in which $\Delta\varphi_{k+1}$ depends only on
$\Delta\varphi_k$, as follows.
In this case
\bea
&p(\Delta\varphi_1,\dots,\Delta\varphi_n)=&
P(\Delta\varphi_{n},\Delta\varphi_{n-1})\dots\nonumber\\
&&\times P(\Delta\varphi_2,\Delta\varphi_1)p_0(\Delta\varphi_1).
\ea{54}
Here $p_0(\Delta\varphi_1)$ is the unconditional probability and
$P(\Delta\varphi_{k},\Delta\varphi_{k-1})$ is a conditional 
probability.
Combining Eqs. \r{54}, \r{53} and \r{int1}, one can obtain that
\be
P_h(n)=\frac{1}{2}+\frac{1}{2}\text{Re}\sum_{\Delta\varphi_{n+1}}
f_{n+1}(\Delta\varphi_{n+1}).
\e{56}
The quantity $f_{n+1}(\Delta\varphi_{n+1})$ is defined by the
following iterative relation, which is convenient for numerical 
calculations,
\bea
&&f_1(\Delta\varphi_1)=p_0(\Delta\varphi_1),\nonumber\\
&&f_{n+1}(\Delta\varphi_{n+1})=\sum_{\Delta\varphi_{n}}
P(\Delta\varphi_{n+1},\Delta\varphi_{n})e^{2i\Delta\varphi_{n}}
f_{n}(\Delta\varphi_{n}).
\ea{55}

The above solution can be conveniently written in a matrix form.
Assuming that the random variable $\Delta\varphi_k$ can assume the
values $\delta\varphi_i$ ($i=1,2,\dots$), we represent  
$p_0(\Delta\varphi_k)$ by a column vector $p_0$ with the components 
$p_0(\delta\varphi_i)$.
The conditional probability is represented by the matrix $P$ with 
$P_{ij}=P(\delta\varphi_i,\delta\varphi_j)$.
Then Eq. \r{56} and \r{55} yield
\be
P_h(n)=\frac{1}{2}+\frac{1}{2}\text{Re}[u(P\Phi)^np_0],
\e{57}
where $u$ is the row vector with the components equal to 1 and $\Phi$
is a diagonal matrix with $\Phi_{jj}=e^{2i\delta\varphi_j}$.

For the non-Markovian model of Sec.\ref{VIB1}
\be
p_0=\frac{1}{2}\left(\begin{array}{c}1\\ 1\end{array}\right),\ \
P=\left(\begin{array}{lr}p&q\\q&p\end{array}\right),\ \
\Phi=\left(\begin{array}{lr}e^{2i\Delta\varphi}&0\\
0&e^{-2i\Delta\varphi}\end{array}\right).
\e{58}


\end{document}